\documentclass{article}

\usepackage[utf8]{inputenc}
\usepackage{url}
\usepackage[english]{babel}
\usepackage{epsfig}
\usepackage{amsfonts}
\title{
	The presheaf of Quantum Realities
	and constructions of the space-time from the Space of Ultimation
}
\usepackage{amsmath}
\usepackage[usenames]{xcolor}

\author{ Innocenti V. Maresin }
\date{December 21, 2012\quad{\small(revised in 2016)}}

\ifx\C\undefined\else\makeatletter\let\C\@undefined\makeatother\fi
\newcommand{\C}{{\mathbb{C}}}
\ifx\R\undefined\newcommand{\R}{{\mathbb{R}}}\fi

\newcommand{\A}{{\mathcal A}}
\newcommand{\Aherm}{{\A_{\mathrm{herm}}}}
\newcommand{\MatRing}[1]{{\mathrm{M}_{#1}(\C)}}

\ifx\Id\undefined\newcommand{\Id}{\mathrm{Id}}\fi
\ifx\Oplus\undefined\newcommand{\Oplus}{\mathop{\oplus}\limits}\fi

\newcommand{\mul}{\hspace{0.125em}}

\newcommand{\conj}[1]{{#1}^*}

\DeclareMathOperator{\cpos}{\circ}

\ifx\im\undefined\DeclareMathOperator{\im}{{\mathrm{im}}}\fi

\newcommand{\UlSp}{{\mathcal E}}
\newcommand{\ball}[1]{{\mathrm{B}_{#1}}}

\newcommand{\QR}{\mathrm{QR}}
\DeclareMathOperator{\Preshv}{{\mathrm{Preshv}}}

\newcommand{\BOp}{{\mathcal B}}
\newcommand{\Hi}{{\mathcal H}}
\newcommand{\pj}{{\mathfrak{p}}}

\newcommand{\Mink}{{\mathbb{M}}}
\newcommand{\Md}{{\Mink^{1+d}}}

\newcommand{\mathhyphen}{{\mathchar"2D}}

\newcommand{\CsOP}{C*alg$^{\mathrm{op}}$}

\begin{document}
\maketitle
\begin{abstract}
The paper presents several quantum models constructed with the formalism of algebraic geometry.
The Universe is presented as a presheaf on the “space of ultimation” (i.e. the branching future)
with values in certain category which enhances the category opposite to C*-algebras.
Given models show the possibility to derive ordinary space-times (of both Special and General Relativity) from ultrametric spaces.
\end{abstract}

\noindent This work is supported in part by the Russian grant NSh-2928.2012.1.\\
This work is supported in part by Russian Foundation for Basic Research,\\grant 10-01-00178.

\subsection{Introduction}
\indent

The paper actually presents two things.
First, it is an interpretation of Quantum Mechanics with multiple possible futures.
Its central concept is a Quantum Reality, which consists of an algebra of observables and {\em one} state on it,
and actually is a variant of the Heisenberg's picture.
This contrasts to the Schr\"odinger's picture, where the state depends on the time coordinate.
Though, Quantum Realities postulate certain non-unitary “causal” evolution of the state, together with the algebra.

\par This is a non-local theory, but not in the same sense as hidden parameters theories are.
Quantum Realities admit that any reality, in some sense, depends on its futures,
and is not related to causally independent realities (except through their common past).

\par Second, an appropriate mathematical formalism is presented,
which is also used to derive the relativistic space-time, from an underlying structure.
This formulation can be used for theories where solutions branch in future directions.
If any reality depends on all its possible futures, then it is handy to parameterize realities
by sets of their futures.
This leads to the concept of the ultimate future, the topological “space of ultimation”
and to application of the presheaf formalism, where “restriction morphisms” to subsets
are, essentially, possible outcomes of a quantum measurement.

\section{Basic concepts}
\subsection{C*-algebras}
\indent

C*-algebra~$\A$ is a complex Banach algebra with an antilinear involution, which reverses the order of multiplication:
$\conj{(a\mul b)} = \conj b\mul\conj a$.
The norm must satisfy:
$$ ||a||^2 = ||\conj a\mul a||$$
The theory of C*-algebras was presented in details, for example, in~\cite{Dixmier}.

$\Aherm$ is a real linear subspace of self-conjugate (Hermitian) elements. 
By definition, the set of “positive” elements $\A^+ = \{\,a\mul\conj a\,|\,a\in\A\,\}$;\footnote{
	An estabilshed, but confusing terminology.
	Actually, it is a generalization for non-negative real numbers $[0,+\infty)$.
} it is a cone in ~$\Aherm$.
\par A *-homomorphism does not have to keep the norm. Though, one can prove that a *-homomorphism cannot increase the norm of an element:
$$ ||f(\conj a\mul a)|| = ||f(a)||^2,\ ||f(\conj a\mul a\mul\conj a\mul a)|| = ||f(a)||^4,
\ ||f((\conj a\mul a)^4)|| = ||f(a)||^8,\ \cdots$$

\par The standard definition of a C*-algebra does not require existence of the unity.
Henceforth, we'll assume that any of our algebras has the “{\bf 1}” element, although it's possible that ${\mathbf 1} = {\mathbf 0}$
(in the \{0\}~algebra, which consists of the only “{\bf 0}” element).
It always exists in a finite-dimensional C*-algebra.
If “{\bf 1}” does not exist in an infinite-dimensional algebra, then it can always be created by adding exactly one dimension.
\par Note that we does not require a *-homomorphism to preserve “{\bf 1}”.
Obviously, $f({\bf 1}) = \conj{f({\bf 1})}\mul f({\bf 1})$ is “positive”.

\subsection{Categories and presheaves}
\indent

{\em Category} is a mathematical concept which formalizes relations between mathematical “objects” (of the same structure)
via {\em morphisms} from one object to another;
refer to~\cite{wiki/Category_(mathematics)} for detailed theory.
It usually, but not always, is used to represent some additional structure on sets.
Most well-known categories (such as sets, groups and other algebraic structures, topological spaces…)
are so named {\em concrete categories},\cite{wiki/Concrete_category}
which means that objects and morphisms can be encoded as sets and functions between them.
Not all categories are concrete, though.
In other words, an object of the category does not necessarily have the structure of a set.

\par The concept of a topological space is assumed to be well-known.
A presheaf~$F$ on a topological space~$X$ with values in a category~C
($ F \in \Preshv(X, \mathrm{C})$)
maps any open subset~$U$ of~$X$ to an object~$F(U)$ of~C.
Also, for any two open subsets $V \subset U \subset X$
a {\em restriction morphism} $F(U)\to F(V)$ must be defined,
which equals to the identity morphism if $V = U$.\\
\parbox[t]{136pt}{
$$ F(W) \hspace{6ex} \to \hspace{6ex} F(U) $$
$$ \to\ F(V)\ \to $$
}
\parbox[t]{200pt}{
The last condition on~$F$ to be a presheaf is for any open $W \subset V \subset U \subset X$ 
the diagram of three restriction morphisms must commute.
This means that the restriction morphism from~$F(U)$ to~$F(W)$
must be equal to the composition of ones from~$F(V)$ to~$F(W)$ and from~$F(U)$ to~$F(V)$.
}
\par Usually, C is a concrete category; then, elements of objects from a presheaf are referred to as germs.
Though, Quantum Realities need the general, categorical definition of a presheaf,
because their morphisms are not functions.

\subsection{Metric spaces, ultrametrics and their generalizations}
\indent

A metric space has a real-valued two-arguments metric function, or distance, such that:
$$d(\cdot,\cdot)\ge 0,
\ \ \forall \epsilon,\zeta:\ d(\epsilon,\zeta)=d(\zeta,\epsilon),
\ \ \forall \epsilon,\zeta:\ (d(\epsilon,\zeta)=0)\Leftrightarrow(\epsilon=\zeta),
$$
$$ \forall \epsilon,\zeta,\eta: d(\epsilon,\zeta)\le d(\epsilon,\eta) + d(\eta,\zeta) $$
The topology of a metric space can be defined through the family of open balls,
$$ \ball r(\epsilon) = \{\,\zeta\,|\,d(\epsilon,\zeta) < r\,\}$$
\par Also, if $$\forall \epsilon,\zeta,\eta: d(\epsilon,\zeta)\le\max(d(\epsilon,\eta),d(\eta,\zeta))\,,$$
then $d$ is referred to as an {\em ultrametric}.\cite{wiki/Ultrametric_space}
One can easily realize that $\varphi\cpos d$,
where $\varphi: [0,+\infty)\to[0,+\infty)$ monotonically increases and $\varphi(0) = 0$,
is also an ultrametric.
This suggests that the codomain of an ultrametric may be any bounded upper semilattice.\cite{wiki/Semilattice}
“Bounded” means that is has “0”, the minimal element.
\footnote{
	For upper semilattices, we'll use notations of calculus (“0” and “max”),
	not the symbol~$\vee$ used in Wikipedia, which is ubiquitous in mathematical logic and algebra.
	The concept of “lattice-valued metric” is seldom used by various authors,
	with this join operation in the triangle inequality and,
	henceforth, we omit the prefix “ultra-”.
}
Though, it is not clear how to make such a space a topological space,
because there is no {\em strict} inequality~“$<$” on a lattice,
which is necessary to define {\em open} balls.
\par For models of the space-time, where the codomain of a (generalized) metric will be referred to
as the Locale of Time, we will assume the following properties of it:
\begin{itemize}
\item It is a bounded upper semilattice with the partial order~“$\le$”,
the minimal element~“0” and the join operation~“max”.
\item It is a non-Hausdorff topological space where for any open~$R$:
$$ \forall \sigma\in R\ \ \forall \tau\le\sigma:\ \tau\in R\,.$$
\item The topology of the Locale of Time is defined by a base, which generalizes intervals~$[0,r)$ of a real-valued ultrametric.
So, the topology of a (generalized) metric space can be defined through the family of open balls
$$ \ball r(\epsilon) = \{\,\zeta\,|\,d(\epsilon,\zeta) \in r\,\},$$
where $r$ belongs to the aforementioned base, which implies that it is an open subset (non-Hausdorff) of the Locale of Time.
\item The intersection of any two subsets in the aforementioned base also belongs to it.
\end{itemize}
\par First two requirements and the ultrametric inequality provide us with the fact that
{\em any} point of an open ball can be its center.
\par For a real-values ultrametric (or, conceptually speaking, if its codomain is totally ordered)
there are no non-trivial intersections of balls:
any two balls either do not intersect at all, or one includes another.
For a generalized metric defined above it is not generally true.
The purpose of the last requirement is to ensure that the intersection of two balls is always a ball.
Since we actually use only balls, but never the value of the metric itself,
the Locale of Time may be even a pointless topological space (also known as {\em locale}),\cite{wiki/Pointless_topology}
where open sets form a special case of a lattice, known as {\em frame}.
\par We'll use almost nothing of a (ultra)metric geometry beyond these two properties.

\section{Quantum Realities and measurements}
\subsection{Categories of quantum realities}

A state~$\rho$ on a C*-algebra~$\A$ is such bounded linear functional that:
$$\forall a\in\A^+: \rho(a)\ge 0\,,\ \ \ ({\bf 1} > 0) \Rightarrow (\rho({\bf 1}) = 1)$$
Remind that ${\bf 1} = 0$ in~\{0\}.
Such functionals corresponds to average values of an observable in quantum mechanics.
If there exist two C*-algebras~$\A_1$ and $\A_2\ne\{0\}$, a *-homomorphism $f: \A_2\to\A_1$
and a state~$\rho_1$ on~$\A_1\,$, then $\rho_1(f({\bf 1}))$ is a non-negative real number.
If it is positive, then
$$
\rho_2 := \frac{\rho_1\cpos f}{\rho_1(f({\bf 1}))}
\mbox{ is a state on } \A_2\,.
$$
\par Let's define the category QR-1 of Quantum Realities.
An object of QR-1 is a C*-algebra~$\A$ and a state~$\rho$ on it.

A morphism from an object $(\A_1,\rho_1)$ to an object $(\A_2,\rho_2)$ is
such *-homomorphism~$f: \A_2\to\A_1$ that 
$$\exists c\in[0,1]\ \ \forall a\in\A_2: \rho_1(f(a)) = c\mul \rho_2(a).$$
We call the factor~$c$ the {\em Born's factor}.
Obviously, $c = \rho_1(f({\bf 1}))$ if $\A_2\ne\{0\}$ and is indeterminate if $\A_2 = \{0\}$.
\par The physical sense of a morphism is a transition (by quantum measurement)
from the reality~1 to the reality~2.
Unless the Born's factor equals to zero, the state functional~$\rho_2$ is determined uniquely,
with aforementioned equation, by the state~$\rho_1$ and *-homomorphism~$f$.
So, unlike observables, states are transformed in the same direction as morphisms in the category QR-1,
from past to future.

\par We'll call the {\em nihil\,\footnote{
	“Nothing” (Latin). Cognates with “annihilation”.
} reality} a reality based on the \{0\}~algebra, and it will be denoted by~(), the 0-tuple.
It is, obviously, the zero object\cite{wiki/Zero_object} of corresponding categories;
its existence is crucial for the presheaf formalism.
It also implies existence of the zero morphism between any pair of realities,
which map all observables to~{0} and have zero Born's factor.

\par It is possible to consider also a full subcategory of QR-1, which includes the nihil object.
For example, the category where all algebras are full matrix algebras (including $\MatRing 0$).

\par Another possibility to define a Quantum Reality category is stateless realities.
The \CsOP\ is the category opposite to the category of C*-algebras.

We'll refer to both QR-1 (and its full subcategories that include “()”)
and \CsOP\ (and its full subcategories that include \{0\})
to as Quantum Reality categories (QR).

\par The {\em singleton} (reality) is the algebra~$\C$ with its only possible state, the identity functional.
It will be denoted by~(1).

Other objects in a category of Quantum Realities based on full matrix algebras and pure states\footnote{
	The state is called {\em pure} if it cannot be represented
	as a non-trivial convex combination of other states.
	For finite-dimensional C*-algebras, any such state has the form
	$\rho(\cdot) = \langle\conj\psi | \cdot | \psi\rangle$, where $\psi$ is a ket-vector, $|\psi| = 1$,
	in the space of the representation of the algebra.
	In full matrix algebras {\em all} states $\langle\conj\psi | \cdot | \psi\rangle$ are pure.
} may be denoted by~$(z_0 : z_1 : \ldots : z_{n-1})$, where “:” means projectivization,
because pure states are actually parameterized by~$\C\mathrm{P}^{n-1}$.

\subsection{Relationship to quantum measurements}
\indent

The most usual type of quantum measurements, a Von Neumann's measurement, relies on the *-homomorphism (and monomorphism)
$$ \BOp(\Hi_l) \to \BOp(\Oplus_k \Hi_k)\, $$
which maps the element~{\bf 1} to a projector and transforms a pure state
$$\psi = \sum\limits_k \psi_k \mbox{\  to \ } \frac{\psi_l}{|\psi_l|^2}
\ (\mbox{if }\psi_l\ne 0)
$$
with Born's factor $|\psi_l|^2$
(it follows from the fact that “{\bf 1}” in~$\BOp(\Hi_l)$ maps to the projector to~$\Hi_l$).
Its simplest example is two homomorphisms from the singleton algebra~$\C$ to the matrix algebra $\MatRing 2$:
$$\pj_{\uparrow}: a \mapsto \begin{pmatrix} a & 0 \\ 0 & 0 \end{pmatrix};
\quad\pj_{\downarrow}: a \mapsto \begin{pmatrix} 0 & 0 \\ 0 & a \end{pmatrix}.$$

\par Another type of measurement represented in the QR formalism
correspond to the natural *-monomorphism of algebras of bounded operators
$$\Oplus_k \BOp(\Hi_k) \to \BOp(\Oplus_k \Hi_k)$$
Note that, since this homomorphism preserves “{\bf 1}”, its Born's factor must be equal to~1.
Several physicists refer to such case as a {\em non-selective measurement};
see e.g. \cite{Ivanov}~5.3.2 or \cite{Pechen}.
\par Its “counterpart” in a complete Von Neumann's measurement is a “selection” homomorphism
$$ \A_l \to \Oplus_k \A_k\,,$$
showing how a mixed reality splits.
Like Von Neumann's measurement, it also maps the element~{\bf 1} to a projector.
Its Born's factor equals to~$\rho({\mathbf 1}_l \oplus 0_{\mathrm{others}})$.
\par Note that Quantum Realities always map observables from the future to the past.
Such things as the “partial trace” has nothing to do with morphisms of Realities.

\subsection{Presheaf formalism and physical restrictions}
\indent

The idea behind this interpretation of Quantum Mechanics
is that any reality, in some sense, includes all its possible futures.
When we advance into the future, the set of available futures shrinks.
Hence, the formalism is based on certain “space of ultimation”~$\UlSp$,
which assumed to have a topological structure.
Its open subsets are called {\em ultimation sets},
and its point are “points of ultimation”, or just “ultimations”.

\par The model of the Universe is described as:
$$ F \in \Preshv(\UlSp, \QR),\ F(\emptyset)=()$$
where $\QR$ is a category of quantum realities (usually, QR-1).
\par From here forth, we will think about realities only as ultimation sets,
and causal relation will be expressed in set-theoretical language.
Namely, $V\subset U$ or $U\supset V$ means “$V$ is a future of~$U$”
or, the same,  “$V$ is a consequence of the cause~$U$”.
It is handy to use inclusion signs to denote the causal relation
instead of arrows (for morphisms in QR),
because it helps to avoid confusion with *-homomorphisms,
whose direction is opposite.

\par The mathematical requirement for~$F$ to be a presheaf is, itself, quite weak.
To construct a physical theory some extra conditions must be satisfied.
The following three requirements, ordered from weakest to strongest,
can be easily guessed from philosophical and quantum-mechanical considerations.

\par Weak {\it ex nihilo nihil fit}\,\footnote{
	“Nothing comes from nothing.” (Latin)
} principle:
\underline{no restriction morphisms from the nihil reality}\\\underline{to a non-nihil reality}.
In other words, a presheaf must be nihil on any subset of a nihil ultimation set.
If we annihilate a presheaf on the union of all nihil ultimation sets
(i.e. put $F = ()$ for any subset of the aforementioned union),
then it will be satisfied.

\par Strong {\it ex nihilo nihil fit} principle:
\underline{any restriction morphism,}\\\underline{but a morphism to the nihil reality, may not be zero}.
It is stronger, because zero restriction morphisms between non-nihil realities are possible in a presheaf.

\par Positive Born's factor principle:
\underline{any restriction morphism,}\\\underline{but a morphism to the nihil reality, must have a positive Born's factor}.
It is stronger than the previous one, because even a non-zero morphism
can map all observables of the future reality to a subalgebra vanished by the state functional of the past reality,
and hence to have the zero Born's factor.
Note that, for a presheaf of stateful quantum realities satisfying this principle,
the state in each reality is uniquely determined by the state
corresponding to the entire Space of Ultrimation (which is also an open set, the Absolute Past),
given algebras and their *-homomorphisms only. 

\par Although this paper does not discuss probabilistic implications,
it should be noted that the positive Born's factor principle
does not preclude “zero-probability” points of ultimation.
In fact, it precludes zero-probability {\em open sets} of ultimation,
which appears to accord with such concept as the spectral decomposition.\cite{wiki/Spectral_decomposition}
It is one of arguments in favor of the presheaf formalism.

\par Henceforth we assume that, in a physical theory of the presheaf on the space of ultimation,
the positive Born's factor principle holds.
Other physically motivated restrictions apparently exist.
For example, from quantum-mechanical theory of measurement
follows that ${\bf 1}$ of $F(U)$ must have a representation
as the sum (possibly, infinite; see e.g. \cite{Dixmier}~B29 for explanations) 
of images of observables from any covering of~$U$ by its ultimation subsets.
Such restrictions are not considered in this paper and will be investigated later.

\subsection{One-dimensional branching time model}
\indent

Suppose that $\UlSp$ is a metric space, with an ordinary, real-valued ultrametric.
If $U = \ball r(\epsilon)$ for certain $\epsilon\in\UlSp$ and $r > 0$,
then there exists a possibility that different $r$ may fit to the same open set $U$.
Namely, let
$$ r_{\mathrm{inf}} := \sup\limits_{\zeta\in U} d(\epsilon,\zeta);
\quad r_{\mathrm{max}} := \inf\limits_{\zeta\in \UlSp\setminus U} d(\epsilon,\zeta);
$$
and if $r_{\mathrm{max}} > r_{\mathrm{inf}}$ then any $r_{\mathrm{max}} \ge r > r_{\mathrm{inf}}$
parametrizes the same ball~$U$.
In such a case, let us think that $U$ corresponds to certain reality
whose duration is the time interval~$[-\log r_{\mathrm{max}}, -\log r_{\mathrm{inf}})$.
If $r_{\mathrm{max}} = r = r_{\mathrm{inf}}$, then the reality $F(U)$ is point-time with $t = -\log r\,$.
If the ball~$V$ is a subset of~$U$, then it's easy to realize that its $r_{\mathrm{max}}$
is not greater than $r_{\mathrm{inf}}$ of~$U$.
Physical interpretation of the restriction morphism of~$F(U)$ to~$F(V)$
is a transition from~$U$ to one of its futures.
Possible future realities or, more strictly,
the covering of~$U$ with smaller balls,\footnote{
	In $p$-adic numbers, and some other examples of ultrametric spaces,
	the radius of balls in the covering can be set as $r_{\mathrm{inf}}$.
	If values of the ultrametric do not form a discrete set even locally
	and $r_{\mathrm{max}} = r_{\mathrm{inf}}$,
	then smaller radii must be chosen for balls of a covering.
}
can form a Von Neumann's quantum measurement,
or other type of it.
\par Note that we have many different realities for any time moment~$t$.
They form a branching structure,
which is equivalent to a directed tree if values of the metric are (locally) discrete
and, hence, $r_{\mathrm{max}} > r_{\mathrm{inf}}$ for any ball.

\pagebreak[3]
\section{Constructions based on the Locale of Time}
\subsection{1+1-dimensional Minkowski space model}
\indent
Suppose that $\UlSp$ is an semilattice-valued metric space,
where the metric takes values in~$[0,+\infty)\times[0,+\infty)$,
the Locale of Time of this particular model.\\
\noindent\parbox[b]{192pt}{\indent

$d_u$ and $d_v$ will denote two real components of this metric,
and open balls (see 1.3.) are defined with:
$$ (d_u < r_u)\wedge(d_v < r_v);\ r_u,r_v \in (0,+\infty)$$
It is, obviously, a semilattice with the componentwise $\max$ operation and $0 = (0,0)$.
Geometrically, it corresponds to the projective 1+1-dimensional Minkowski space,
more strictly, to the conformal compactification of the 1+1d Minkowski space.
Namely, the (temporal) infinity~$\Omega$ maps to~0,
and two light infinities map to $\{0\}\times(0,+\infty)$ and $(0,+\infty)\times\{0\}$ respectively.
}\epsffile{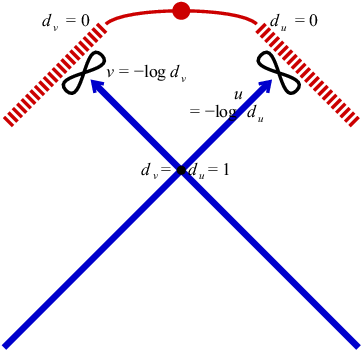}\\
On the other hand, the affine part (the Minkowski space proper) 
map to $(0,+\infty)\times(0,+\infty)$.

\par The main difference of this model from the model of~2.4.\ is that
this Locale of Time and radii of balls defined with it are only partially ordered.
In one-dimensional time, all ultimation balls between some large (past) ball and some smaller (future) ball
are totally ordered by their radii.
A semilattice-valued metric makes a more complicated causal structure,
which, in this case, resembles the graph product of two directed trees,
a directed acyclic graph, but not a tree itself.

\subsection{The EPR paradox}
\indent

This subsection gives an interpretation of the quantum entanglement
and Einstein–Podolsky–Rosen paradox
within the model constructed in the previous subsection.
It does not give an explicit construction of the space of ultimation and the presheaf,
but concrete examples may be created from this description.

\par Suppose we have a metric ball~$B$
where the algebra of observables of~$F(B)$
has the form $\MatRing 4 \otimes \mathrm{Environment}$
and the state has the form $\psi \otimes \rho_{\mathrm{Environment}}$,
where $\psi$ is some pure state~$(\psi_{\uparrow\uparrow}:\psi_{\uparrow\downarrow}:\psi_{\downarrow\uparrow}:\psi_{\downarrow\downarrow})$.
\par Define four *-homomorphisms from~$\MatRing 2$ to $\MatRing 4$
and four *-homomorphisms from~$\C$ to $\MatRing 4$ (see 2.2.):
$$\pj_{\uparrow\mathhyphen}: \begin{pmatrix} a_1 & a_2 \\ a_3 & a_4 \end{pmatrix} \mapsto
\begin{pmatrix} a_1 & 0 & a_2 & 0 \\ 0&0&0&0 \\ a_3 & 0 & a_4 & 0 \\ 0&0&0&0 \end{pmatrix};
\ \ \pj_{\downarrow\mathhyphen}: \begin{pmatrix} a_1 & a_2 \\ a_3 & a_4 \end{pmatrix} \mapsto
\begin{pmatrix} 0&0&0&0 \\ 0 & a_1 & 0 & a_2 \\ 0&0&0&0 \\ 0 & a_3 & 0 & a_4 \end{pmatrix};
$$
$$\pj_{\mathhyphen\uparrow}: {\mathbf a}\mapsto\begin{pmatrix} {\mathbf a} & 0 \\ 0 & 0 \end{pmatrix};
\ \ \pj_{\mathhyphen\downarrow}: {\mathbf a}\mapsto\begin{pmatrix} 0 & 0 \\ 0 & {\mathbf a} \end{pmatrix};
\ \ \pj_{\alpha\beta} := \pj_{\mathhyphen\beta}\cpos\pj_{\alpha} = \pj_{\alpha\mathhyphen}\cpos\pj_{\beta}\,.
$$
\par Let an “advance by~$u$” to denote a transition to a smaller ball
with smaller~$d_u$ (the first component of the two-component ultrametric) and the same~$d_v\,$,
and, similarly, an “advance by~$v$” to denote a transition to a smaller ball with smaller~$d_v\,$ and the same~$d_u\,$.
Suppose that a sequence of one on more advances by~$v$ from~$B$ changed only the “Environment”.
It means that the algebra~$\A'_v$ of a $v$-advanced reality~$F(B'_v)$
has the same structure $\MatRing 4 \otimes \mathrm{Environment}'_v\,$,
the morphism from~$F(B)$ to~$F(B'_v)$ correspond to a *-homomorphism from~$\A'_v$ to~$\A$
of the form $\mathrm{id}_{4\times 4} \otimes E'_v\,$.
On the next step, let the ball split to such two $v$-advanced balls $B_{\mathhyphen\uparrow}$ and $B_{\mathhyphen\downarrow}$,
where $\A_{\,\mathhyphen\beta} = \MatRing 2 \otimes \mathrm{Environment}_{\,\mathhyphen\beta}\,$,
that *-homomorphisms from~$\A_{\,\mathhyphen\beta}$ to~$\A$ have the form $\pj_{\mathhyphen\beta} \otimes E_{\mathhyphen\beta}\,$.
\par Similarly, suppose the same picture on~$u$
and a split to two $u$-advanced balls $B_{\uparrow\mathhyphen}$ and $B_{\downarrow\mathhyphen}$,
that *-homomorphisms from~$\A_{\alpha\mathhyphen}$ to~$\A$ have the form $\pj_{\alpha\mathhyphen} \otimes E_{\alpha\mathhyphen}\,$.
\par What can we say about ultimation intersections $B_{\alpha\beta} := B_{\alpha\mathhyphen}\cap B_{\mathhyphen\beta}\,$,
namely, $B_{\uparrow\uparrow}, B_{\uparrow\downarrow}, B_{\downarrow\uparrow}$,
and $B_{\downarrow\downarrow}$?
First of all, some intersections may be empty.
We supposed that $\UlSp$ has a structure like the product of two metric spaces,
but it is not necessarily such a product,
and the intersection of a $v$-ball with $u$-ball can be empty.
\\
\noindent\parbox[t]{240pt}{
But what corresponding algebras of observables appear to be?
Because morphisms between Realities commute,
their images~$\im\A_{\alpha\beta}$ in~$\A$
must lie in the intersection $\im\A_{\alpha\mathhyphen}\cap\im\A_{\,\mathhyphen\beta}$
of images of observables from corresponding $u$-advanced and $v$-advanced realities.

\par These intersections have the form {\small
$\begin{pmatrix} 1 & 0 & 0 & 0 \\ 0 & 0 & 0 & 0 \\ 0 & 0 & 0 & 0 \\ 0 & 0 & 0 & 0 \end{pmatrix}
\otimes$} some subalgebra of Environment, for~$\uparrow\uparrow$, and so on.
If for the original~$\psi: \psi_{\uparrow\downarrow} = \psi_{\downarrow\uparrow} = 0$
(with $\psi_{\uparrow\uparrow} = \psi_{\downarrow\downarrow}$ it gives an EPR~pair),
then balls $B_{\uparrow\downarrow}$ and $B_{\downarrow\uparrow}$ must be nihil,
because corresponding Born's factors must be 0.
A nihil intersection can be either the empty set, or a non-empty set annihilated by the presheaf.
}\hspace{8pt}\parbox[t]{124pt}{ \colorbox{green}{
\begin{tabular}{c@{}c@{}c@{}c@{}c}
$\scriptstyle\begin{pmatrix}\psi_{\uparrow\uparrow}\\\cdot\\\cdot\\\cdot\end{pmatrix}$
& \colorbox{white}{$\displaystyle \psi_{\uparrow\uparrow}$} & \colorbox{yellow}{!}
& \colorbox{magenta}{$\displaystyle \psi_{\uparrow\downarrow}$}
&
$\scriptstyle\begin{pmatrix}\cdot\\\psi_{\uparrow\downarrow}\\\cdot\\\cdot\end{pmatrix}$
\\
& \colorbox{cyan}{- - -} & {}
& \colorbox{cyan}{- - -}
\\
$\scriptstyle\begin{pmatrix}\cdot\\\cdot\\\psi_{\downarrow\uparrow}\\\cdot\end{pmatrix}$
& \colorbox{magenta}{$\displaystyle \psi_{\downarrow\uparrow}$}
& \colorbox{yellow}{!}
& \colorbox{white}{$\displaystyle \psi_{\downarrow\downarrow}$} &
$\scriptstyle\begin{pmatrix}\cdot\\\cdot\\\cdot\\\psi_{\downarrow\downarrow}\end{pmatrix}$
\\
\end{tabular}
}
\\\vspace{-6pt}
\epsffile{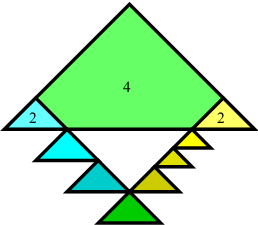}
}
\par This example demonstrates how Quantum Realities approach the problem of locality.
Although $v$-advanced realities do not depend on $u$-advanced realities and vice versa,
they not always can combine to a joint future,
which makes an illusion that a measurement in a $v$-advanced reality
can change something in a $u$-advanced reality.

\pagebreak[3] 
\subsection{Locale of Time in more dimensions}
\indent

The main obstacle to apply this construction straightforwardly to 1+$d$-dimensional Minkowski space $\Md$
is that pseudo-Riemannian manifolds for $d > 1$ are not (semi)lattices.
It means that the Locale of Time cannot be made from $\Md$ proper,
with only addition of some infinite points.
The Locale of Time can be constructed,
although a detailed construction (for $\Md$ and some other globally hyperbolic manifolds,
as well for space-times with different geometries in different futures)
has to be presented in separate papers.\footnote{
	One construction of the Locale of Time is presented in 2016 in \cite{CRFfLM}.
}
A possible way to generalize the Locale of Time construction correctly
is to consider the conformal compactification of~$\Md$ and its infinite part~$\mathcal{I}$.

\section{Miscellaneous considerations}
\subsection{The choice of direction of morphisms and direction of time}
\indent

Yes, both are firmly based on physical grounds.
A (pre)sheaf of C*-algebras cannot encode a quantum measurement
in such a way that possible results are parts of the original reality.
There exist the *-homomorphism from “measured” to “original”,
but an opposite direction is impossible, al least for full matrix algebras,
because “measured” has less dimensions than “original” (if we understand a measurement as a decomposition).
\par The direction of time is determined, first, by semantics of the “measurement”.
Also, a morphism of Quantum Realities easily maps an original pure state to a future mixed state, but not versa.
It's this which appears to be compatible with entropy considerations.

\subsection{Relationship to algebraic quantum field theory}
\indent

The algebraic quantum field theory\cite{wiki/Algebraic_quantum_field_theory}
also uses sheaves and C*-algebras,
but there are two major differences between it and the Quantum Realities.
First, in algebraic quantum field theory the direction of morphisms is not reversed.
This allows it to use products (see the next subsection), but restriction morphisms in Algebraic QFT are not measurements.

Second, algebraic quantum field theory postulates the space-time, not derives it.


\subsection{Why not sheaves?}
\vspace{-0.5ex}\indent
 
The definition of a sheaf\cite{wiki/Sheaf_(mathematics)} requires the product of objects,
but categories of Quantum Realities with states have no such operation.
The category \CsOP\ has the product~– it is the coproduct of C*-algebras,
i.e. the free product completed in the appropriate norm,
an infinite-dimensional algebra for any two non-nihil factors.
Though, it unlikely has a physical sense.

Let us prove that any Quantum Reality category with states cannot have the product operation.
By contradiction, suppose that there exists the product~$(1)\times(1)$ of two singletons.
By definition,\cite{wiki/Product_(category_theory)}
for an object~$S$ any pair of morphisms $m_1: S\to (1),\ m_2: S\to (1)$
such morphism $m: S\to (1)\times(1)$ must exist that:
$$ m_k = \pi_k\cpos m\,,\mbox{ where }\pi_k\mbox{ are canonical projections.}$$
Due to symmetry (and a universal property of the product) $\pi_k$ must have the same Born's factor
and it is some definite number, because $(1)\times(1)$ cannot be nihil.
This implies that $m_1$ and $m_2$ always have equal Born's factors,
which contradicts to the requirement that they are {\em any} pair of morphisms.


\subsection{Relationship to categorical quantum mechanics}
\vspace{-0.5ex}\indent

Another application of formalism of the theory of categories to quantum physics
is categorical quantum mechanics.\cite{wiki/Categorical_quantum_mechanics}
But that formalism describes quantum {\em systems}; there are operations on systems
such as the composite system (tensor product).
The formalism of Realities can consider (sub)systems
only inside an algebra of observables (as we demonstrated in~3.2.),
but not on the categorical level.
There is no such operation as composite system in the Quantum Realities formalism.
Such operation cannot be used with the presheaf formalism
because a composite system of two systems having the common past
would violate the no-cloning principle.
\par In short, Quantum Realities are not Categorical QM in any way.

\subsection{Origin of symmetries}
\vspace{-0.5ex}\indent

Ultrametric constructions do not explain the origin of physical symmetries,
such as translations, Lorentz transforms and gauge groups.
It's plausible that symmetries do not rely on the space of ultimation,
but are properties of algebras of observables,
where corresponding (families of) vector fields are preserved by homomorphisms between realities.
Some insights can be found in the paper~\cite{VectFi} of the same author.
If such vector fields generate semigroups,
then homomorphisms should be intertwiners of corresponding endomorphism semigroups in different realities.

\subsection{Is the space of ultimation actually metric?}
\vspace{-0.5ex}\indent

First, I think that it is a physical object.
Second, I do not think that it has not, in fact, a disjoint topological structure,
because it makes all branches “sharp”,
which is not accepted well by the physical intuition.
Possibly, the space of ultimation is not even a Hausdorff space.
Ultrametric spaces are, indeed, a good approximation
of the actual structure of the space of ultimation.
Ultrametric spaces were used for many years 
in researches of conformational spaces of large, complicated molecules (such as proteins),
but without realizing its theoretical value
in application to the process of branching of the time.
Certainly, a conformational space is not an ultrametric space.
But it should be approximated with an ultrametric space,
because its terrible complexity hinders the use of more accurate models.
Actually, the conformational space of a molecule
which occupied certain state in a given moment of the past,
is a projection of the corresponding ultimation set.



\end{document}